\documentclass{mn2e}
\usepackage{times}
\usepackage{bm}
\usepackage{graphicx}

\title[Disk accretion]{Disk accretion onto a magnetized star}
\author[Ya.~N. Istomin, P.~Haensel]{Ya.~N. Istomin $^{1}$, P.~Haensel $^{2}$ \\
${1}$ P.N.~Lebedev Physical
Institute, Leninsky Prospect 53, Moscow 119991, Russia, E-mail:
istomin@lpi.ru, \\
${2}$ N. Copernicus Astronomical Center, Polish
Academy of Sciences, Bartycka 18, PL-00-716 Warszawa, Poland, E-mail: haensel@camk.edu.pl}
\begin{document}
\date{}
\pagerange{\pageref{firstpage}--\pageref{lastpage}} \pubyear{2012}
\maketitle
\label{firstpage}
\begin{abstract}
The problem of interaction of the rotating magnetic field, frozen to a star,
with a thin well conducting accretion disk is solved exactly. It is shown that a disk
pushes the magnetic field lines towards a star,
compressing the stellar dipole magnetic field.
At the point of corotation, where the Keplerian rotation frequency coincides
with the frequency of the stellar rotation, the loop of the electric current appears.
The electric currents flow in the magnetosphere
only along two particular magnetic surfaces, which connect the corotation
region and the inner
edge of a disk with the stellar surface. It is shown that the closed current surface
encloses the magnetosphere. Rotation of
a disk is  stopped at some distance from the stellar surface,
which is 0.55 of the corotation radius. Accretion from a disk spins up
the stellar rotation. The angular momentum transferred to the star is determined.
\end{abstract}
\section{Introduction}
The problem of interaction of an accretion disk  with a magnetized star is very
important for understanding of compact X-ray sources. The energy release for a
variable X-ray sources and X-ray pulsars is due to the accretion of the matter
from a companion star onto the surface of a compact star, namely, on a neutron star.
Falling onto a neutron star, a particle gains the energy per unit mass equal to
the value of the gravitational potential on the stellar surface $\varphi=0.15 c^2 (M_s/M_\odot)
(R_s/10 km)^{-1}$. where $M_s$ and $R_s$ are stellar mass and radius, respectively,
and $M_\odot$ is the solar mass. The energy release
depends on how the accretion is realized. The magnetic field, which is frozen to
a neutron star, is high enough to prevent a free falling of the matter on a
star. The naive point of view is that the accretion is free outside the so called
Alfv\'en surface, and then continues along the magnetic field lines down to the polar
caps of a star (Lamb 1989). In a standard approach Alfv\'en surface is the surface where the magnetic field pressure
$B^2/8\pi$ is equal to the pressure of the accreting matter. But here it is not
clear what is the pressure of the accreting gas. Is it
the total dynamic pressure $\varrho v^2/2$ or its radial part only, $\varrho v_r^2/2$~?
The Alfv\'en radius is estimated for the spherical accretion, $v=v_r$, but then
is applied to the disk accretion. But in this case
the gas velocity $v$ comes mainly from the velocity of disk rotation and is of the
order of Keplerian velocity $v_K$, which is much larger than the radial
velocity. The gradient of the magnetic pressure in the azimuthal direction
is small and therefore does not oppose significantly the disk rotation. Besides it is clear that
the disk must compress the magnetic field within the disk inner edge. These features
do not help to answer the question: where is the real position of the Alfv\'en surface.

An accreting gas is a plasma consisting of ions and electrons,
which have a quite different Larmor radii in a magnetic field.
Moving across the magnetic field, the plasma gets polarized and
creates the large scale electric field ${\bf E}$. If this field is
${\bf E}=-{\bf v}\times{\bf B} /c$ then the plasma continues to
drift in the crossed ${\bf B, E}$ fields at the same velocity
${\bf v}$ not slowing down. The real value of ${\bf E}$ depends on
the boundary conditions on a disk and a star surface, on electric
currents in a plasma and on conditions of their closing. The
important characteristics are the electric conductivity of a disk
plasma and the neutron star conductivity in the surface layers.
The ionized plasma of an accretion disk has the conductivity
$\sigma = 10^{13}(T_e/1 eV)^{3/2}(\Lambda/10)^{-1}\quad s^{-1}$,
which is high enough to consider a disk as an ideal conductor.
Here $T_e$ is the temperature of electrons in a disk, which is
higher than $10 eV$; $\Lambda$ is the Coulomb logarithm
($\Lambda\simeq 20$). At such a high conductivity $\sigma$ the
width of the skin layer $\lambda_{sk} = (\tau c^2/\sigma)^{1/2}$
is less than the disk width $H$. The value of $\tau$ is the
characteristic time of turbulent motion in the $\alpha$-disk,
$\tau = H/v_K$. The condition $\sigma\gg c^2/H v_K$ is well
fulfilled in the inner parts of a disk. As for the conductivity of
surface layers of a neutron star $\sigma_{NS}$, then it is as high
as $ 10^{21} s^{-1} $ (Blandford et al.
 1983; Potekhin 1999). Therefore, $\sigma_{NS}
\gg \sigma$, and a neutron star can be considered as ideal conductor too.

As ideal conductor, a disk tends to exclude the stellar magnetic field, pushing it
toward a star. Heavy disk ions tend to rotate with the Keplerian velocity, but magnetized
electrons are frozen to the magnetic field lines, which rotate with the angular
velocity of a star $\omega_s$. Thus, the point $\rho=\rho_c$ where a disk corotates
together with a star, $\omega_s\rho_c=(G M_s/\rho_c)^{1/2}$,
$$
\rho_c=\left(\frac{G M_s}{\omega_s^2}\right)^{1/3} = 1.5\cdot 10^8\left(
\frac{M_s}{M_\odot}\right)^{1/3}\left(\frac{P_s}{1s}\right)^{2/3} cm,
$$
is the point beginning from which ($\rho<\rho_c$) motions of ions and electrons
differ essentially. Here $G$ is the gravitational constant
and $P_s=2\pi/\omega_s$ is the period of the stellar rotation.
The region of corotation is namely the region where the interaction of an
accreting disk with the stellar magnetic field begins. The point of corotation
is inside the stellar light cylinder radius $R_L=c/\omega_s$ for all rotating
neutron stars: $P_s>3\cdot 10^{-5}(M_s/M_\odot) s$. On the other hand, the
corotation radius is larger than the neutron star radius $R_s$. Because of that
we will consider the undisturbed magnetic field of a neutron star as dipolar. Also we
assume that the axis of a dipole is parallel to the neutron star rotation axis.

The paper is organized as follows. In section 2 we describe an arbitrary axisymmetric
magnetic field. In section 3 we find the motion of ions in a disk. Section 4 describes
electric currents in the stellar magnetosphere. In section 5 we find the structure of the
magnetic field. Finally, in Section 6 we find the torque acting on a star and
discuss the obtained results.

\section{Axisymmetric Magnetic Field}

Since our aim is to find the magnetic field of the magnetosphere of a star,
we initially introduce convenient variables, which describe it more simply.
For an axisymmetric magnetosphere, when the angle between the axis of the
magnetic dipole and axis of the rotation of a star equals zero, it is useful
to introduce the flux of the poloidal magnetic
field
$f(\rho,z)$. Here $\rho$ and $z$ are the cylindrical coordinates. Then the
components of the magnetic field are
\begin{equation}
B_\rho= -\frac{1}{\rho}\frac{\partial f}{\partial z}, \quad B_z=
\frac{1}{\rho}\frac{\partial f}{\partial\rho}, \quad B_\phi =\frac{1}{\rho}g \quad .
\end{equation}
The function $g(\rho,z)$ describes the toroidal magnetic field produced by the
poloidal electric currents flowing in a stellar magnetosphere. The relation
$f=constant$ is the equation for the magnetic surface on which the magnetic
field lines are lying. It is also
convenient to describe magnetic field lines of the poloidal magnetic field
not only as $f=f(\rho,z)$, but also $\rho=\rho(z,f)$. The poloidal
magnetic field is
$$
B_\rho = \frac{1}{\rho}\left(\frac{\partial\rho}{\partial z}\right)_f
\left(\frac{\partial\rho}{\partial f}\right)_z^{-1}; \quad
B_z = \frac{1}{\rho}\left(\frac{\partial\rho}{\partial f}\right)_z^{-1}.
$$
For the dipole magnetic field the magnetic flux $f(\rho,z)$ is
$$
f_d = \frac{B_sR_s^3\rho^2}{(\rho^2+z^2)^{3/2}},
$$
where $B_s$ is the magnitude of the
magnetic field on the surface of a star on its equator,
$B_s=|B_z(r=R_s)|_{eq}$.
For our problem it is important to know the value of magnetic flux in the
equatorial plane, $z=0,\, f_0(\rho)=f(\rho,z=0)$. Then the vertical magnetic
field on the equator is $B_z(z=0)=\rho^{-1}\partial f_0/\partial\rho$.

It is useful to measure the magnetic field in the units of $B_s$ and the
distances
$\rho,z$ in terms of star radius $R_s$; then the units for the magnetic
flux
will be $B_sR_s^2$. The relations (1) are the same in these dimensionless
variables. For a dipole field, the dimensionless flux is $f_d^\prime =
{\rho^\prime}^2/({\rho^\prime}^2+{z^\prime}^2)^{3/2},\, {f_0^\prime}_d=
1/\rho^\prime$.

In the star magnetosphere, electric currents flow along the magnetic field
lines, ${\bf j}=a{\bf B}$, where $a({\bf r})$ is an arbitrary scalar. In this
case, the poloidal components of the Maxwell equation, $\nabla\times{\bf B}=
4\pi{\bf j}/c$, result in the relations
$$
g(\rho,z)=g(f),\quad a=\frac{c}{4\pi}\frac{{\rm d}g}{{\rm d}f}.
$$
These relations mean that the toroidal magnetic field and the electric currents
in an axisymmetric magnetosphere are functions of the poloidal magnetic flux,
$f$. The important case is the absence of toroidal volume electric currents,
$j_\phi=0$. Then $(\nabla\times{\bf B})_\phi =0$, and the equation for the
magnetic flux $f$ is the Laplace type
\begin{equation}
\rho\frac{\partial}{\partial\rho}\left(\frac{1}{\rho}\frac{\partial f}{\partial
\rho}\right)+\frac{\partial^2 f}{\partial z^2}=0.
\end{equation}
The function $f$ is neither even nor odd with respect to the coordinate $z$ due to
the electric currents flowing in accretion disk in the equatorial plane. Thus,
we expand the function $f$ over the exponential functions $\exp(-\lambda z)$,
for $z>0$. The solution of the equation (2) is
\begin{equation}
f(\rho,z)=\int_0^\infty\exp(-\lambda z)\lambda\rho J_1(\lambda\rho)\varphi
(\lambda)d\lambda;\quad z>0,
\end{equation}
where $J_1$ is Bessel function of the first kind. An arbitrary function
$\varphi(\lambda)$ in this integral is defined by boundary conditions for the
equation (2).
For the dipole magnetic field function $\varphi$ is $\varphi(\lambda)=1$
(in dimensionless units). The deviation of $\varphi$ from the unity describes
the
distortion of the stellar magnetospheric magnetic field from dipolar. It is useful to
express the function $\varphi$ through the value of the magnetic flux on the
boundary $z=0,\, f_0(\rho)$. Because of $f_0(\rho)=\int_0^\infty\lambda
\rho J_1(\lambda\rho)\phi(\lambda)d\lambda$,
using the inverse transformation, we get
\begin{equation}
\varphi(\lambda)=\int_0^\infty f_0(\rho^\prime)J_1(\lambda\rho^\prime)d
\rho^\prime.
\end{equation}
 The equations (3) and (4) determine the magnetic field through the boundary
condition $f_0(\rho)$. Of course, we also take into account another
boundaries: infinity and the axis $\rho=0$. The quantity $f(\rho,z)$ is
finite there except the point $\rho^2+z^2\rightarrow 0$, where
$f\rightarrow f_d$.

\section{The Accretion Disk}
The disk is produced from the matter falling from a companion star to a neutron star.
The rate of accretion is of ${\dot M}$. The centrifugal force prevents a disk
matter from spreading over a magnetosphere, and this gives rise to an
equatorial plasma disk. The
width of this disk is small compared with its radial size. The strong magnetic
field of a neutron star forces
the disk plasma to rotate with the angular velocity of a star.
We will describe motions of ions and electrons separately because their  motions are quite
different in the strong magnetic field of a star. Disk ions have
radial, $v_\rho$,
and azimuthal velocities, $v_\phi$. We consider a stationary and axisymmetric
star magnetosphere. Due to axial symmetry, all quantities are independent
of the azimuthal angle $\phi$.
Then, the continuity equation for ions is
\begin{equation}
\frac{1}{\rho}\frac{\partial}{\partial \rho}(\rho n_i v_\rho)=\frac{{\dot M}}
{2\pi\rho m_i}\delta(\rho - \rho_{out})\delta(z),
\end{equation}
where $n_i$ is ion density, $\rho_{out}$ is the
outer radius of the disk, $\rho_{out}\gg\rho_c$. Integrating  Equation (5) over
$\rho$ and $z$, we obtain
\begin{equation}
\Sigma v_\rho=-\frac{{\dot M}}{2\pi\rho m_i}\theta(\rho_{out}-\rho).
\end{equation}
Here $\Sigma$ is the surface disk density, $\Sigma=\int n_i dz$.
Let us note the very important fact that the divergence of the ion flux
at $\rho<\rho_{out}$ equals zero.

Now we write the equation of motion of ions
\begin{equation}
m_i({\bf v}\cdot\nabla){\bf v}=q_i{\bf E}+\frac{q_i}{c}{\bf v}\times {\bf B} + {\bf F}.
\end{equation}
The star acts on ions by gravitational force, $F_\rho =-m_i G M_s/\rho^2$,
which, neglecting other forces,
must result in the rotation of ions with the Keplerian velocity, $v_K=(GM_s/
\rho)^{1/2}$.
The $\rho$-component of Equation (7) gives
\begin{equation}
v_\rho\frac{\partial v_\rho}{\partial \rho} -\frac{v_\phi^2}{\rho}=
\frac{q_i}{m_i}\left( E_\rho+\frac{v_\phi}{c}B_z\right)-\frac{v_K^2}{\rho}.
\end{equation}
We can see from Equation (8) that in the absence of the electromagnetic
fields $B_z$ and $E_\rho$, the ions rotate with the Keplerian velocity,
$v_\phi = v_K$. In contrast, in a strong magnetic field $B_z$, ions have the
electric drift velocity, $v_\phi = -cE_\rho/B_z$. For corotation, i.e. the rigid rotation
with the angular velocity of the star, $v_\phi=\omega_s\rho$, the
plasma in the disk must be polarized
to create the radial electric field, $E_\rho = -\omega_s\rho B_z/c$. The corotation
velocity is less than the Keplerian velocity at the distance $\rho<\rho_c$.
Because the Keplerian velocity goes down with increasing distance $\rho$, the corotation
of a plasma disk means that the centrifugal force in the disk becomes less
than the gravitation
force at $\rho<\rho_c$, and ions must move inwards in the radial direction.
But this is not easy, due to the conservation of
angular momentum. That is described by the $\phi$-component of the equation (7),
\begin{equation}
v_\rho\frac{\partial v_\phi}{\partial \rho} + \frac{v_\rho v_\phi}{\rho}=-
\frac{q_i}{cm_i}v_\rho B_z.
\end{equation}
Let us note that there is no the toroidal electric field, $E_\phi=0$, because
under stationary conditions the electric field must be potential.
Because the ion radial velocity, $v_\rho$, is not equal to zero
(see  Equation (6)), the solution of Equation (9) is
\begin{equation}
\rho v_\phi +\frac{q_i}{cm_i}f_0(\rho)=const.
\end{equation}
Equation (10) is the conservation law of the total angular momentum of an
ion in magnetic field. But the dimesionless coefficient before the second term in
Equation (10), written as $(q_iB/cm_i\omega_s)(\omega_sf_0/B)$, is much greater than unity.
It is proportional to the ratio of
the ion cyclotron frequency of rotation in the magnetic field of a
star, $\omega_{ci}=q_iB_s/cm_i$, to the frequency of rotation of star,
$\omega_s$. We denote this ratio $\Omega_c=\omega_{ci}/\omega_s$. This is
the cyclotron frequency of ions
in natural units of the stellar rotation frequency
$$ \Omega_c=1.5\cdot 10^{12}\left(\frac{P_s}{1 sec}\right)\left(\frac{B_s}{10^9 G
}\right).
$$
For the
dipole
magnetic field, the terms in the left hand side of Equation (10) become
of the same order only at large distances $\rho/R_s\simeq\Omega_c^{1/3}$.
Angular momentum conservation prohibits ion radial
motion in such a strong magnetic field as that of a star if we assume it
is (close to) dipolar one. The only possibility to allow ions to move radially
is an expulsion of magnetic field from the disk. Only if the magnetic flux changes slowly in
the disk, $\Delta f_0\simeq B_s\rho^2\Omega_c^{-1}$, does the motion of
ions in radial direction become
possible. But a small variation of the magnetic flux $f_0$ with the radial
distance $\rho$ means a small value of the vertical magnetic field $B_z$ in
the equatorial plane (see formula (1)).

As for electrons, the parameter $\Omega_c$ for them is
at least $10^3$ times greater than for ions. The plasma electrons are
strongly magnetized and can move only along a magnetic surface $f=const$.
They generate a current stream at $f=f_c=f_0(\rho=\rho_c)$ towards a star.
Thus, the disk plasma is separated at $\rho<\rho_c$: ions create the plasma disk in the
equatorial plane, but electrons create the electric current $J_c$ on the magnetic surface
$f=f_c$. This does not mean that the plasma disk is positively charged, it is neutral.
The electrons from the 'sea'
of electrons of a stellar magnetosphere neutralize any electric charge.

\section{Magnetospheric and Star Surface Currents}

We consider the magnetospheric plasma which has an infinite conductivity.
Under stationary conditions it can rotate around the z axis with the
angular velocity $\omega(\rho,z)$. In the frame rotating with this velocity
the electric field ${\bf E}'$ is equal to zero. So, the electric field in
the laboratory frame is
$$
{\bf E}=-\frac{1}{c}(\bm{\omega}\times{\bf r})\times{\bf B};\,
E_\rho=-\frac{\omega}{c}\frac{\partial f}
{\partial \rho},\, E_z=-\frac{\omega}{c}\frac{\partial f}{\partial z},\,
E_\phi =0.
$$
The electric field must be potential ${\bf E}=-\nabla \Psi$, where $\Psi$
is the electric potential. This means that the frequency of rotation
$\omega(\rho,z)$ is only function of the magnetic flux $f,\, \omega=
\omega(f)$, and the electric potential is $\Psi=\int^f\omega
(f^\prime)df^\prime/c$. Because the conductivity of a star surface is very
high, $\omega(f)=\omega_s$.

The magnetospheric current ${\bf J}_c$ flows from a star surface to a disk along the
magnetic surface $f=f_c$. This
current is closed by the star surface current ${\bf J}_s$.
The surface current, ${\bf J}_s$, flowing on the stellar surface is connected
with the volume magnetospheric current, ${\bf j}$, by the continuity
equation
$$
\nabla_s\cdot {\bf J}_s = -j_n.
$$
The divergence in this equation is taken along the surface where surface
currents flow, and $j_n$ is the normal (to this surface) component of the
volume current ${\bf j}$. The analogous equation connects the ion surface
current of the disk ${\bf J}_d$ with the magnetospheric volume current
${\bf j}$. But the
divergence of the ion surface current, as we have already emphasized (see equation (6)),
equals zero. The electrons of the disk can only rotate with the angular
velocity of the magnetosphere $\omega(f)$, and the divergence of their flux is
also equal zero. Thus, the volume magnetospheric current ${\bf j}$ is
absent
except on the particular magnetic surfaces $f=f_c$ and $f=f_s$. The value of
$f=f_s$ is the magnetic flux where the disk stops rotation, $f_s>f_c$.
The total current $I$ produced by the accretion flow is
\begin{equation}
I=\frac{q_i{\dot M}}{m_i}=6\cdot 10^{20}\left(\frac{{\dot M}}{10^{-10} M_\odot/y}
\right) A.
\end{equation}
One half of it flows in the north hemisphere and another half flows in
the south hemisphere. The magnetospheric currents flow along the magnetic
field on two magnetic surfaces
\begin{equation}
{\bf j}_s = -{\bf B}\frac{q_i{\dot M}}{4\pi m_i}\delta(f-f_s), \quad
{\bf j}_c = {\bf B}\frac{q_i{\dot M}}{4\pi m_i}\delta(f-f_c).
\end{equation}
These currents are closed by the surface current flowing in the stellar polar region.

The magnetospheric electric currents ${\bf j}_s, {\bf j}_c$ produce in the
region $f_s>f>f_c$ the toroidal magnetic field $B_\phi$,
\begin{equation}
B_\phi=\frac{g}{\rho},\quad g=\frac{q_i{\dot M}}{c m_i}.
\end{equation}
Then the $\phi$-component of the Maxwell equation
$\nabla\times{\bf B}=4\pi{\bf j}/c$ gives the equation for the poloidal
magnetic flux
\begin{equation}
\rho\frac{\partial}{\partial\rho}\left(\frac{1}{\rho}\frac{\partial f}{\partial
\rho}\right)+\frac{\partial^2 f}{\partial z^2}=\frac{g^2}{2}[\delta(f-f_s)-
\delta(f-f_c)].
\end{equation}
The right hand side of the equation (14) is not equal to zero only on the
magnetic surfaces $f=f_s, f_c$, where the magnetospheric electric currents
flow along the magnetic field. Due to the toroidal component of the magnetic
field the longitudinal currents have a toroidal component, which distorts the
poloidal magnetic field. Outside the surfaces $f_s, f_c$, Equation (14)
coincides with Equation (2) and its solution is given by relations (3,4).
However, the fields on both sides of the discontinuities are different. The
connection between the magnetic fields at $f>f_s$ and $f<f_s$, and also between
$f>f_c$ and $f<f_c$, can be obtained by integrating the equation (14) over
$f$ in the small region near surfaces $f=f_s$ and $f=f_c$ respectively. For
that we need to go over to another independent variables, $(z,f)$, and consider
the magnetic surfaces to be determined by the relation $\rho=\rho(z,f)$.
Equation (14) becomes
\begin{eqnarray}
\frac{1}{2}\left[1+\left(\frac{\partial\rho}{\partial z}\right)^2\right]
\frac{\partial}{\partial f}\left[\left(\frac{\partial\rho}{\partial f}\right)^{-2}\right]+
\left(\frac{\partial\rho}{\partial f}\right)^{-2}\frac{\partial}{\partial f}
\left[\left(\frac{\partial \rho}{\partial z}
\right)^2-\right. \nonumber \\
\left.\ln\left|\frac{\partial\rho}{\partial z}\right|\right]-
\left(\frac{\partial\rho}{\partial f}\right)^{-1}\left(\frac{2}{\rho}+
\frac{\partial^2\rho}{\partial z^2}\right)= \frac{g^2}{2}[\delta(f-f_s)-\delta(f-f_c)].
\end{eqnarray}
We take into account that on both sides of discontinuity the shapes of the magnetic
surfaces are the same
$$
\left.\frac{\partial\rho}{\partial z}\right|_{f=f_{s,c}+0}=\left.
\frac{\partial\rho}{\partial z}\right|_{f=f_{s,c}-0}.
$$
Then integrating the equation (15) we obtain
\begin{equation}
B_\rho^2+B_z^2+B_\phi^2=const.
\end{equation}
This is the natural condition of equality of the magnetic pressures on both
sides of a discontinuity, which follows exactly from our equations.

\section{Structure of Magnetic Field }

Here and below we will use the dimensionless variables
\begin{eqnarray}
&\omega^\prime=\omega/\omega_s,\,{\bf r}^\prime={\bf r}/R_s,\,{\bf v}^\prime=
{\bf v}/R_s\omega_s,\,{\bf B}^\prime={\bf B}/B_s,\nonumber \\
&f^\prime=f/B_sR_s^2,\,
\Sigma^\prime=2\pi m_i\omega_s R_s^2\Sigma/{\dot M},\, {\bf J}^\prime=4\pi
R_s{\bf J}/I.
\end{eqnarray}
Further on we will omit the index $\prime$. There appear several
dimensionless parameters. The ratio
of the ion cyclotron frequency to the frequency of star rotation, mentioned alrready in
the section 3, is
\begin{equation}
\Omega_c=\frac{q_iB_s}{cm_i\omega_s}.
\end{equation}
The second parameter is the magnetic field, created by the current $I/2$
(11) flowing on the star surface, in the units of $B_s$
\begin{equation}
{\cal B}=\frac{q_i{\dot M}}{c m_i R_sB_s},
\end{equation}
$$
{\cal B}=60\left(\frac{{\dot M}}{10^{-10}M_\odot/y}\right)
\left(\frac{B_s}{10^{12}G}\right)^{-1}\left(\frac{R_s}{10^6cm}\right)^{-1}.
$$

Moving toward a star a disk pulls out the star magnetic field
lines. Vertical component of the magnetic field $B_z$ decreases
and the ions can move in the radial direction. But the electrons
remain magnetized. They must move along magnetic field surfaces
only. The single possibility for them to change their motion from
the Keplerian rotation to the corotation with a star is to do that
at the point of corotation $\rho=\rho_c$. Thus, a disk banks all
dipole magnetic field lines crossing the equatorial plane at
$\rho>\rho_c$ into the region $\rho<\rho_c$. This must had
happened because magnetic field lines were froze initially to the
ideal conducting matter of a disk and had to move together with
it. We conclue therefore that the stellar magnetosphere under the
action of an ideal accretion disk is confined to the region
$\rho<\rho_c$! Thus, the magnetic field flux at the point
$\rho=\rho_c$ equals zero, $f_c=0$. The magnetic field line $f=0$
just comes from the magnetic pole of a star. The topology of
magnetic field line is drawn on the figure 1.

%%%%%%%%%%%

\begin{figure}
\begin{center}
\includegraphics[height=10cm]{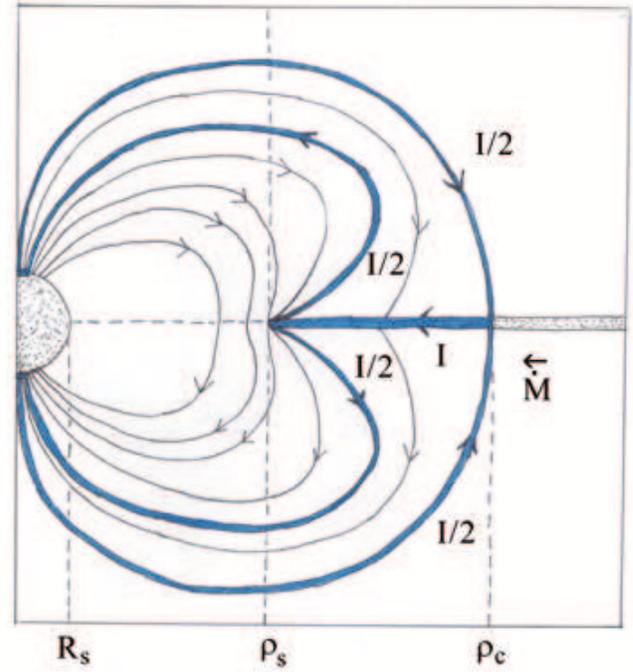}
\end{center}
\caption{Topology of the magnetic field and currents. Thick lines are electric currents.}
\label{fig:Bdisk}
\end{figure}

%%%%%%%%%%%%%

The velocity of rotation of plasma disk ions is given by Eq. (10)
\begin{equation}
v_\phi=\frac{\rho_c^2}{\rho}-\frac{\Omega_c f_0(\rho)}{\rho}.
\end{equation}
Here we consider that at the point $\rho=\rho_c$ the velocity
of ion rotation is equal to the Keplerian velocity $v_K$. Eq. (20) fixes
the value of magnetic flux at the stopping of the disk rotation $v_\phi=0$,
$f_s=\rho_c^2/\Omega_c$.
The electrons of the disk rotate with the angular velocity of rotation
of the magnetic field lines $\omega(f)=\omega_s$, however.
Thus, the plasma disk produces the surface toroidal electric current
\begin{equation}
J_\phi=2{\cal B}\Sigma(v_\phi-\rho),
\end{equation}
which gives the discontinuity of the radial magnetic field on the surface
$z=0$.
\begin{equation}
\left. B_\rho\right|_{z=z+0}=-\left. B_\rho\right|_{z=z-0}={\cal B}\Sigma
(v_\phi-\rho).
\end{equation}
The value of $(v_\phi-\rho)$ is a function of $\rho$, $f_0(\rho)$
and is determined by the relations (20)
\begin{equation}
v_\phi-\rho=\frac{\rho_c^2}{\rho}-\frac{\Omega_c f_0}{\rho}-\rho.
\end{equation}
It is seen from (21,23) that the value of the radial magnetic field up to the
stopping point $\rho=\rho_s$ is negative, $B_\rho(\rho, z=+0)<0$.
At the point $\rho=\rho_c$ the radial magnetic field in the
equatorial plane is equal zero, and the magnetic field line here is perpendicular
to a disk. Then magnetic field lines bend toward a star.

The value of the radial magnetic field $B_\rho(\rho,z=z+0)$ determines the
derivative of the magnetic flux over $z$ on the equator
\begin{equation}
\left.\frac{\partial f}{\partial z}\right|_{z=+0}=-{\cal B}\Sigma\rho
(v_\phi-\rho).
\end{equation}
Let us find this derivative as a function of $f_0(\rho)$. From the equations
(3,4) we obtain
$$
\left.\frac{\partial f_0}{\partial z}\right|_{z=+0}=-\rho\int_0^\infty d\rho^\prime f_0(\rho^\prime)\int_0^\infty \lambda^2 J_1(\lambda\rho)J_1(\lambda\rho^\prime)d\lambda.
$$
The second integral is known (Oberhettinger 1972), and we get
\begin{eqnarray}
\left.\frac{\partial f_0}{\partial z}\right|_{z=+0}=\frac{2}{\pi}\rho\frac{\partial}
{\partial\rho}\frac{1}{\rho}
\left\{\int_0^1\frac{dxK(x)}{dx}f_0(\rho/x)dx- \right.\nonumber \\
\left.\int_0^1\frac{dK(x)}{dx}f_0(\rho x)dx\right\}.
\end{eqnarray}
Here $K(x)$ is the complete elliptic integral of the first kind. It equals
$\pi/2$ at $x=0$, but has the logarithmic singularity at $x\to 1$, $K(x)\to
\ln[16/(1-x^2)]/2$. Equation (25) must be supplemented with the condition of absence of the radial magnetic
field in the region $\rho<\rho_s$, $\partial f_0/\partial z=0$. Let us note that substitution
of the dipole magnetic field, $f_0=1/\rho$, into Eq. (25) satisfies this condition for all $\rho$.
Also the square of the vertical magnetic field $B_z^2$ behind the stopping point, $\rho=\rho_s-0$, must be equal to the sum of the square of the
the radial one $B_\rho^2$ ($B_z<<B_\rho$) and the square of the toroidal one $B_\phi^2={\cal B}^2/\rho_s^2$ before the stopping point
(see the condition (16)). As a result we get $B_z=-{\cal B}(1+\rho_s^2/v_\rho^2(\rho=\rho_s))^{1/2}/\rho_s$. We see that the magnetic field is
strongly compressed here by a disk, $|B_z|>>|B_{dip}|=\rho_s^{-3}$.
The equations (24-26) close the system of equations which
determine the function $f_0(\rho)$. We add also the energy conservation law
following from the equations (8,9)
\begin{equation}
\frac{1}{2}(v_\rho^2+v_\phi^2)-\frac{\rho_c^3}{\rho}+\Omega_cf_0(\rho)=const(\rho).
\end{equation}
The particle energy decreases due to the work of the radial electric field
in the disk on ions. This electric field is generated in the disk by the
rotating magnetic field. Considering that the radial disk velocity $v_\rho$
is smaller than the rotation velocity $v_\phi=v_K$ at the point $\rho=\rho_c$,
we find the radial velocity of a disk at any point $\rho_s<\rho<\rho_c$
$$
v_\rho^2(\rho)=\frac{2\rho_c^3}{\rho}-\rho_c^2-v_\phi^2-2\Omega_cf_0(\rho).
$$
The radial velocity $v_\rho$ defines the surface density $\Sigma(\rho)=
1/(\rho|v_\rho|)$. The radial velocity $v_\rho$ at the stopping point $\rho=\rho_s$
does not equal zero and is negative
$$
v_\rho(\rho=\rho_s)=-\rho_c\left[2\left(\frac{\rho_c}{\rho_s}-\frac{3}{2}\right)\right]^{1/2}.
$$
At the point $\rho=\rho_s$ disk ions begin to move along
the magnetic surface $f=f_s$ toward the star surface to its polar cap.
Particles without the mechanical angular momentum ($v_\phi=0$) fall down to a star freely but along curved
fixed magnetic lines. They get the kinetic energy $v^2/2=\rho_c^3-\rho_c^2$ on the stellar surface
which corresponds
to fall down from the height of the corotation point $\rho_c$.

The function $f_0(\rho)$ on all axis $0<\rho<\infty$ is different in three
regions: $f_0=f_0^{(1)}(\rho)$ at $\rho<\rho_s$; $f_0=f_0^{(2)}(\rho)$ at
$\rho_s<\rho<\rho_c$; $f_0=0$ at $\rho>\rho_c$. Thus, the equation (25) becomes
\begin{eqnarray}
&\frac{\partial}{\partial\rho}\frac{1}{\rho}\left\{\int_{\rho/\rho_s}^1\frac{dxK(x)}{dx}
f_0^{(1)}(\rho/x)dx + \right. \nonumber \\
&\left.\int_{\rho/\rho_c}^{\rho/\rho_s}\frac{dxK(x)}{dx}f_0^{(2)}(\rho/x)dx\right\}=0, \, \rho<\rho_s.
\end{eqnarray}
\begin{eqnarray}
&\frac{2}{\pi}\rho\frac{\partial}{\partial\rho}\frac{1}{\rho}
\left\{\int_0^{\rho_s/\rho}\frac{dK(x)}{dx}f_0^{(1)}(\rho x)dx+
\int_{\rho_s/\rho}^1\frac{dK(x)}{dx}f_0^{(2)}(\rho x)dx\right\} \nonumber \\
&=\frac{{\cal B}}{|v_\rho|} \left(\frac{\rho_c^2}{\rho}-
\frac{\Omega_c f_0^{(2)}}{\rho}-\rho\right), \, \rho_s<\rho<\rho_c.
\end{eqnarray}
The value of the magnetic flux $f_0^{(2)}$ is small. At the point $\rho=\rho_s$ it equals $f_0^{(2)}=f_s<<1/\rho_s$, then it decreases to zero at $\rho=\rho_c$.
On the contrary, the function $f_0^{(1)}(\rho)$ increases strongly from the value $f_0^{(1)}=f_s$ at
$\rho=\rho_s$ to the dipole one $1/\rho_s$ at $\rho<\rho_s$ due to the large derivative $\partial f_0^{(1)}/\partial\rho
=-{\cal B}(1+\rho_s^2/v_\rho^2)^{1/2}$. The width of this transition region $\Delta\rho$ is small,
$\Delta\rho\simeq 1/(\rho_s{\cal B})<<\rho_s$. Here the magnetic field is large, it is
strongly compressed by the disk. Then, approaching to a star, the magnetic field becomes
dipolar $f_0^{(1)}=1/\rho$. The function $f_0(\rho)$ is plotted in
the figure 2.

%%%%%%%%%%

\begin{figure}
\begin{center}
\includegraphics[height=8cm]{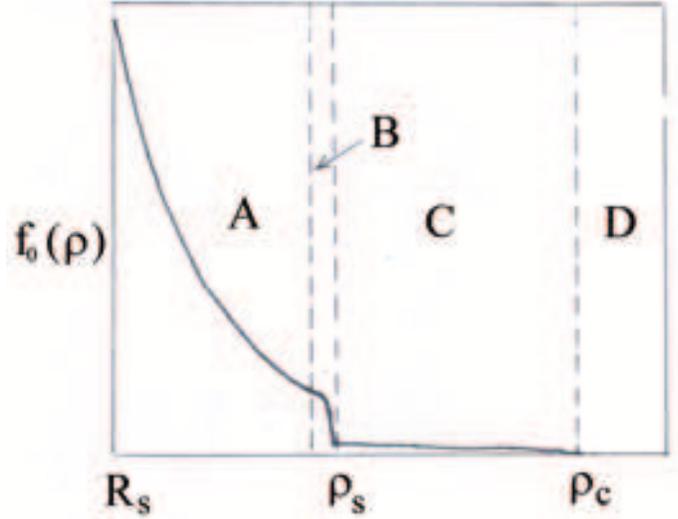}
\end{center}
\caption{Dimentionless poloidal magnetic flux in the equatorial plane, $f_0$,
versus distance $\rho$ from the z axis. Four regimes of the $\rho$-dependance are
distinguished:\, {\bf A}\, $R_s<\rho<\rho_s-\Delta\rho$ ($\Delta\rho<<\rho_s$) where
the dipolar formula is valid, $f_0(\rho)=1/\rho$;\,\, {\bf B}\, Narrow region
$\rho_s-\Delta\rho<\rho<\rho_s$ where the poloidal field lines are strongly
compressed by the disk, and $f_0(\rho)$ decreases rapidly from $\approx 1/\rho_s$
down to $f_s\ll 1/\rho_s$;\, {\bf C}\, $\rho_s<\rho<\rho_c$. Here, $f_0(\rho)$ is small
and decreases from $f_s$ to $f_0(\rho_c)=0$. \,{\bf D} \, For $\rho>\rho_c$, $f_0(\rho)=0$. For
details see section 5.}
\label{fig:f0rho}
\end{figure}

%%%%%%%%%%%%%%%%%%%%%

The position of the stopping point $\rho_s$ can be determined from the equation (28). At the corotation
point $\rho_c$ the right hand side in (28) equals zero. The main contribution to the l.h.s. comes from the dipole
part of $f_0$, $f_0^{(1)}=1/\rho$. Substituting this dependence, integrating, differentiating and
equating the result to zero at point $\rho_c$, we obtain the equation
$$
E\left(\frac{\rho_s}{\rho_c}\right)=K\left(\frac{\rho_s}{\rho_c}\right)\left(1-\frac{\rho_s^2}
{\rho_c^2}\right).
$$
Here $E(x)$ is the complete elliptic function of the second kind. The root of this equation,
except the trivial one,
$\rho_s=0$, is $\rho_s=0.55\rho_c$. The point $\rho=\rho_s$ can be called the Alfv\'en point. But it
corresponds to
the stopping of the disk rotation, not of the radial motion. And its position is fixed by the corotation
radius and does not depend on the star magnetic field $B_s$. That is due to the strong compression of
the dipole magnetic field by a disk in the region $\rho<\rho_s$.

\section{Discussion}

We have shown that the structure of the rotating neutron star magnetic field, interacting
with a thin accretion disk, can be exactly
solved, and we obtained an analytical solution for this field. To achieve this we have
used ideal approximations: an axisymmetric,
well conducting magnetosphere with a thin plasma disk. These
conditions are adequate to the reality. We have demonstrated that a disk
compresses the stellar magnetic field, pushing it towards a star.
The magnetic field is compressed in the region $\rho<\rho_s$,
where $\rho_s$ is the inner edge of the disk where it stops rotation in the laboratory frame, $\rho_s=0.55\rho_c$.
After that the matter of a disk, with zero mechanical angular momentum, falls down
onto a star along the magnetic field surface $f=f_s$.
The loop of the electric current starts near the point of corotation $\rho=
\rho_c$. The electric current $I$ (11) flows along the disk at $\rho_s<\rho<
\rho_c$, then half of it, $I/2$, flows along the magnetic surface $f=f_s$ to the stellar surface , then
along the stellar surface in the polar region, and finally returns to a disk
along the magnetic surface $f=f_c=0$. There is a symmetric current structure in the second hemisphere.
Let us emphasize that the magnetosphere exists only inside the magnetic surface $f=f_c=0$. It is maintained
by the pinching Amp\'ere force $f_a={\bf j}_e\times {\bf B}_\phi/c$ acting on the electron current ${\bf j}_e$ flowing
outside a star along the magnetic
surface $f_c$. This force just balances the gradient of the magnetic pressure $\nabla B^2/8\pi$ inside the magnetosphere boundary.

The potential difference $\Psi$ created
by the rotating magnetic field is (Landau et al. 1984)
$$
\Psi =\frac{\omega_s}{c}f_s.
$$
The electric current $I$, flowing along the star surface perpendicular the
magnetic field, creates the Amp\'ere force and spins up the star rotation.
The torque $K$ is
\begin{equation}
K=I\Psi/\omega_s={\dot M}\rho_c^2\omega_s.
\end{equation}
It produces the acceleration of the star rotation ${\dot P}_s$
\begin{eqnarray}
{\dot P}_s =&-1.5\cdot 10^{-13}\left(\frac{P_s}{1 s}\right)^{7/3}\left(\frac{{\dot M}}{10^{-10}M_\odot/y}
\right) \nonumber  \\
&\left(\frac{M_s}{M_\odot}\right)^{2/3}\left(\frac{I_s}{10^{45}g cm^2}
\right)^{-1}. \nonumber
\end{eqnarray}
Here $I_s$ is the moment inertia of a star. This value of ${\dot P}_s$ corresponds to the
observed one for the close X-ray binaries (Rappaport \& Joss 1983).
We see that the total angular momentum, which has a disk at the point of corotation,
is transferred to a star. The answer does not depend on the strength of the stellar magnetic
field $B_s$. It is due to the strong magnetization of a disk plasma. The magnetization
parameter $\Omega_c$ (18) is much greater than unity. And no matter whether it is very large ($10^{12}$),
or not so much ($10^{5}$). The role of the stellar magnetic field consists in stopping of the disk
rotation, i.e. transformation of the mechanical angular momentum of particles to the electromagnetic
momentum. The angular momentum of a charged particle in a magnetic field consists of two parts:
the mechanical one and the electromagnetic one (see the formula (10)). And approaching a star,
a particle loses its mechanical part and increases the electromagnetic one. Therefore, if a matter
element falls down onto a star, ${\dot M}>0$, then it transfers its angular momentum to a star, $K>0$.
No matter whether it is the mechanical angular momentum or the electromagnetic one.
Thus, under the accretion, there must not be
the negative part in  the torque, which can be attributed to the
electromagnetic torque (Ghosh et al. 1977).
Appearance of the stellar spin down is connected with assumptions that, first, the toroidal magnetic field $B_\phi$
changes its direction at some value of the radius, and, second, the value of the poloidal field $B_z$ in the disk does
not differ strongly from the dipole one (for example, see Kluzniak \& Rappaport
2007). The toroidal magnetic field
is determined by poloidal currents, and the zero value of it at some distance means the closure of currents.
At larger distance the field $B_\phi$ is equal to zero. Indeed, we see that the toroidal field exists only in the
region $\rho_s<\rho<\rho_c$ and $B_\phi>0$ (13).

Let us note that at the small value of accretion rate ${\dot M}$ it is impossible to construct the self
consistent solution for the stellar magnetosphere. When the magnetic field at the point $\rho_s$ becomes
less than the dipole one at this  point it means that there appears the decompression of the magnetic
field. This is impossible for the plasma moving toward a star because the magnetic field is frozen to
the well conducting matter. Formally, in this case it is impossible to match the solution for
the magnetic
flux $f_0(\rho)$ with the dipole field. This takes the place when $|B_z(\rho=\rho_s)|<\rho_s^{-3}$.
Because $|B_z|={\cal B}(1+\rho_s^2/v_\rho^2)^{1/2}/\rho_s$,
$\rho_s=0.55\rho_c$ and $v_\rho=-0.8\rho_c$, this condition is ${\cal B}<2\rho_c^{-2}$. It means that the expression (29)
for the torque is valid when ${\dot M}>{\dot M}_{min}$,
\begin{eqnarray}
&&{\dot M}_{min}=2\frac{cm_i}{q_i}\frac{B_sR_s^3}{\rho_c^2}= \nonumber  \\
&&3\cdot 10^{-16}\left(\frac{\mu_s}{10^{30}G cm^3}
\right)\left(\frac{\rho_c}{10^8 cm}\right)^{-2}\, M_\odot/y.
\end{eqnarray}
Here $\mu_s$ is the stellar magnetic moment, $\mu_s=B_sR_s^3$. For ${\dot M}<{\dot M}_{min}$ a disk does
not penetrate into the star magnetosphere.

In some sense, the stellar magnetic field does not prevent the accretion of a ionized matter onto a star,
but, on the contrary, helps it. Without magnetic field an ideal matter element can not accrete due to the angular
momentum conservation. We need to introduce an anomalous viscosity represented by $\alpha$ parameter
in the $\alpha$-disks. In a magnetic field a charged particle
possesses two parts of the angular momentum, the mechanical one and the electromagnetic one. In the
language of magneto hydrodynamics,
we have the viscous stress tensor and the magnetic stress tensor, which is proportional to the product
$B_\phi B_z$. Appearance of the toroidal magnetic field is impossible without poloidal electric currents,
they must be closed on the stellar surface, transforming the mechanical angular momentum of a disk into the stellar angular momentum.

It is clear that if the inner edge of a disk $\rho_d$ is placed
further than the the corotation point, then it will extract the angular momentum from a star. In this case the
centrifugal force is larger than the gravitation force, and a disk must be pushed outward from a star. This is
the propeller regime. It is stationary, or quasi stationary process, if we have a source of the matter at the inner
edge of a disk. Such situation takes place in the Jupiter magnetosphere where the Jupiter's satellite Io
supplies a  gas to a plasma disk at the point $\rho_d>\rho_c$. The electric current in the current
loop has the direction opposite to that in the case of accretion (Istomin 2005). If there is no gas source
in the stellar magnetosphere then a disk is ejected out to the light cylinder taking the angular momentum $\Delta J$
from a star of the order of
$$
\Delta J\simeq M_d R_L^2\omega_s.
$$
Here $M_d$ is the mass of a disk and $R_L$ is the radius of the
light cylinder, $R_L=c/\omega_s$.

Up to now we considered only the situation when a disk and a star rotate in the same direction
(corotating disk).
There exists the interesting case when an accretion disk rotates in the opposite direction than
a star. A corotating disk had
the same sign of the angular momentum as a star. The disk matter falling onto star transmits its angular
momentum to a star. Then, the torque is positive. If the disk rotates in the opposite direction
than a star (counterrotating disk), then, matter
falling onto the stellar surface, will spin down the stellar rotation. The torque becomes negative.
For the countercotating
disk it is impossible to build up the stationary self consistent solution like we did for the coratation
case. A star must destroy the inner parts of a disk approaching too close to it. It tends to re-rotate the inner part of a
disk to the direction of its own rotation. In the other words, the magnetosphere-disk coupling
tends to enforce corotation to the inner layers of the disk.
The accretion becomes more quasispherical than disk-like. We can estimate with a good
accuracy the torque acting on a star in this case. And the exact solution obtained above will help us. We saw that to reduce
to zero its mechanical angular momentum, a falling particle must deflect from its position on
the magnetic field surface $f$ by
the value of $\Delta f=\rho_m v_m/\Omega_c$ (see formula (20)). Here values $\rho_m$ and $v_m$ are
the radial position and the azimuthal velocity of a particle on the magnetospheric boundary,
respectively. For the rotating disk, the boundary of the magnetosphere is the corotation
radius, $\rho=\rho_c$. For the counterrotation there is no a corotation point. The magnetospheric
boundary for the quasi spherical accretion must be close to the Alfv\'en radius, $\rho_m\simeq\rho_A$.
Falling onto a star, ions move almost along the magnetic field lines deflecting by a small value
of $\Delta f$. Electrons will
deflect also, but their magnetization parameter $\Omega_c$ is much larger than that for ions, and we can neglect electron deflection.
Thus, ions and electrons will be on the stellar surface at different magnetic surfaces separated by the value of $\Delta f$.
The current $I$ (11) will flow along the surface, it will produce the torque $K$ (29). For the
counterrotation of a star and a disk
the value of $\Delta f$ will be negative, and the electric current on the stellar surface will have
the opposite direction than for
the case of corotation. As a result the torque acting on a star will be negative and equal to $K=-{\dot M}v_m\rho_m$.
That is the exact relation. But the position of the magnetosphere boundary is not known exactly. If we do not take into account
the magnetic field compression on the boundary it is given by the traditional formula for the Alfv\'en radius
$\rho_m=\rho_A=(\mu_s^2/2^{1/2}{\dot M}\rho_c^{3/2}\omega_s)^{2/7}$ (see,
for example, Arons 1993).
Because for the accretion
onto a star the radius of the magnetosphere $\rho_m$ must be smaller than the corotation radius $\rho_c$,
the deceleration torque will be always less than the acceleration torque at the same value of the accretion rate ${\dot M}$.
Thus, the observed stellar spin down with noticeable ${\dot M}$ can be a result of the accretion
of the matter from a counter rotating disk.
This is quite possible because the disk formation depends on a companion  star, on the stellar orbital motion and many other reasons.

In conclusion we compare the results obtained here with the numerical calculations. The
numerical simulations were
done in the frame of the MHD approximation by Romanova et al. 2011, 2012. Though our treatment have been done in
the frame of two-fluid
hydrodynamics, the similar features were seen in numerical single-fluid simulations: the compression of the stellar dipole magnetic
field by a disk, the falling of the disk matter along fixed magnetic field lines, the pushing out
of magnetic field lines from a disk,
and the proportionality of the torque $K$, acting on a star, to the accretion mass rate ${\dot M}$.

\section*{Acknowledgements}

Authors thank V.S. Beskin and J.L. Zdunik for the helpful discussions.
The work was supported by the Polish MNiSW research grant no. N N203 512838.

\end{document}